\documentclass[a4paper,10pt]{article}

\usepackage[utf8]{inputenc}
\usepackage[USenglish]{babel}
\usepackage{color,soul}
\usepackage{amssymb}
\usepackage{amsmath}
\usepackage{amsthm}
\usepackage{cite}
\usepackage{graphicx}
\usepackage[space]{grffile} 
\usepackage{dcolumn}
\usepackage{bm}
\usepackage{afterpage}
\usepackage{xcolor}
\usepackage{wasysym}
\usepackage{epstopdf}
\usepackage{hyperref}
\usepackage{comment}
\usepackage{dirtytalk}
\usepackage[affil-it]{authblk}
\usepackage{lipsum}
\usepackage[symbol]{footmisc}
\usepackage[labelfont=bf]{caption}
\usepackage[normalem]{ulem} 
\usepackage{setspace}
\usepackage{geometry}

\hypersetup{
	colorlinks=true,
	linkcolor=blue,
	filecolor=magenta,      
	urlcolor=blue,
	citecolor=red,
}
\newgeometry{vmargin={20mm}, hmargin={20mm,20mm}} 
\newcommand{\vs}{\emph{versus~}}
\newcommand{\arr}{\leftrightarrow}
\newcommand{\larr}{\leftarrow}
\newcommand{\rarr}{\rightarrow}
\newcommand{\la}{\langle}
\newcommand{\ra}{\rangle}
\newcommand{\TO}{\!\to\!}
\newcommand{\GE}{\!\ge\!}
\newcommand{\LE}{\!\le\!}
\newcommand{\GG}{\!\gg\!}
\newcommand{\LL}{\!\ll\!}
\newcommand{\APPROX}{\!\approx\!}
\newcommand{\+}{\!+\!}
\renewcommand{\-}{\!-\!}

\newcommand{\IBM}{individual-based model }
\newcommand{\MF}{mean-field }
\newcommand{\kappac}{\kappa_\mathrm{c}}
\newcommand{\kappap}{\kappa_\mathrm{p}}
\newcommand{\kappamax}{\kappa^\mathrm{max}}
\newcommand{\tf}{t_\mathrm{f}}
\newcommand{\x}{\boldsymbol{x}}

\begin{document}

\title{Influence of invasion on natural selection in dispersal-structured populations}

\author[1,2]{D. {Navidad Maeso}}
\author[1]{M. Patriarca}
\author[1]{E. Heinsalu\thanks{\emph{e-mail}: els.heinsalu@kbfi.ee (E. Heinsalu).}}
\affil[1]{\small{National Institute of Chemical Physics and Biophysics - Akadeemia Tee 23, Tallinn 12618, Estonia}}
\affil[2]{\small{Tallinn University, School of Natural Sciences and Health - Narva 29, 10120 Tallinn, Estonia}}

\maketitle

\begin{abstract}
We investigate how initial and boundary conditions influence the competition dynamics and outcome in dispersal-structured populations.
The study is carried out through numerical modeling of the heterogeneous Brownian bugs model, in which individuals are characterized by diversified diffusion coefficients and compete for resources within a finite interaction range.
We observed that if instead of being distributed randomly across the domain the population is initially localized within a small region, the dynamics and the stationary state of the system differ significantly.
The outcome of the natural selection is determined by different competition processes, fluctuations, and spreading of the organisms.
\end{abstract}




\section{Introduction}

The motion of organisms is together with demographic processes a fundamental feature of life, being a crucial component in evolution and ecology.
There is a variety of ways how organisms move and a wide set of corresponding causes, ranging from escaping predation and decreasing competition to foraging and searching for a mate or a suitable habitat.
The general understanding of animal movement in all its forms, from dispersal to animal migration and swarming, is an extremely challenging task~\cite{Holden-2006}.
A relevant problem is the influence of the dispersal of individuals on the outcome of the competition~\cite{Okubo-Levin, Lewis-book, McPeek-1992}, which has been investigated for a long time using various mathematical models~\cite{May77, Comins_etal_80, Levin-1984, Kessler-2009, Pigolotti-2014-PRL, Pigolotti2016, Waddell-2010, Novak-2014, Johnson-1990, Lin-2015a,Hastings-1983, Holt-1985, Dockery_etal_98, Hutson-2003, Dieckmann-1999, Hutson-2001, Baskett-2007}. 
The results of such investigations show that while noise and temporal fluctuations give a competitive advantage to faster species~\cite{Pigolotti-2014-PRL, Kessler-2009, Waddell-2010, Novak-2014, Johnson-1990, Lin-2015a}, disorder or heterogeneity e.g. in the form of inhomogeneous distribution of nutrients favor slower individuals~\cite{Hastings-1983, Holt-1985, Dockery_etal_98, Hutson-2003, Dieckmann-1999, Hutson-2001, Baskett-2007}.

In previous works~\cite{HNMP-2020,HNMP-2018} we studied natural selection in a dispersal-structured population of Brownian bugs, in which organisms (bugs), initially distributed randomly across the whole domain and with diversified diffusion coefficients, perform a random walk and undergo a Verhulst population dynamics with finite-range competition.
It was observed that the outcome of the competition process cannot be interpreted simply in terms of a natural selection of the most or least motile individuals, as in Refs.~\cite{Pigolotti-2014-PRL, HHL-2013}.
Instead, the competitive advantage is strongly conditioned by pattern formation in the system and can be given to individuals with different --- small, intermediate, or high --- diffusion coefficients, depending on the system parameters~\cite{HNMP-2018, HNMP-2020}.

In this paper we continue and extend the previous study to the ecologically relevant case of a population initially localized in a small fraction of the domain.
Examples of systems that could be described by such a model are a heterogeneous population of micro-organisms initially concentrated within a small area and diffusing across a region with available resources~\cite{Gloria2008,Dennehy-Livdahl-2004} or the colonization process of an initially empty island or region carried out by a small heterogeneous population~\cite{macarthur2001}.
As we will show, the spatio-temporal evolution of the system turns out to be a complex process characterized by the appearance of a wave front dominated by faster species propagating in the empty part of the domain that it is followed by the spreading of other species.
The corresponding outcome of natural selection leads to stationary diffusivity distributions different from those corresponding to a population initially distributed randomly across the whole domain.

%

The problem studied in this paper is related to that of evolution on a wave front, where faster species are selected and the evolution process leads to peculiar phenomena such as front acceleration~\cite{Hallatscheketal_2007,Champagnat-2007, Arnold-2012a, Benichou_etal_2012, Hallatschek-2014a, Hudson-2016a}. 
Investigating the underlying natural selection process can in turn be useful for the understanding of evolution on a wave front in the presence of boundaries associated with the finite size of the system.
We also discuss the dynamics of the bugs model in relation to the question of the dispersal-competition trade-off and its role in species coexistence~\cite{Cadotteetal_2006,Nadell-Bassler-2011,Larocheetal_2016}.

The structure of the paper is as follows.
In Sec.~\ref{Sec-system} we summarize the heterogeneous bugs model investigated.
In Sec.~\ref{Sec_num_exps} we study the selection process taking place in a heterogeneous set of bugs expanding from the center of a domain under different geometrical configurations. 
We characterize the spreading of the bugs by studying the wave front velocity and show that it is described by the Ganan-Kessler velocity formula~\cite{Ganan-kessler-2018}.
We outline and discuss the results obtained and possible future research lines in Sec.~\ref{Sec-conclusion}.

\section{The dispersal-structured Brownian bugs model} 
\label{Sec-system}

We study the Brownian bugs model that describes organisms that reproduce asexually, die and move in space according to Brownian diffusion. 
We assume that there are $N_0$ bugs present in the system at the initial time $t=0$ and that each bug $j$ is characterized by a different diffusion coefficients $\kappa_j$ ($j = 1, \dots, N_0$) extracted randomly from a uniform distribution in the interval $[0, 2 \kappa]$, with mean value $\mu = \kappa$ and standard deviation $\sigma = \kappa /\sqrt{3}$.
The form of the uniform distribution implies that mean and standard deviation are proportional to each other, $\sigma \propto \mu$, so the larger is the mean value $\kappa$, the larger is the variation of the diffusivity among the individuals.
In Ref.~\cite{HNMP-2018} a more general distribution, in which mean and standard deviation are dependent parameters, was considered, namely a uniform distribution with diffusivities in the range $((1-d)\kappa,(1+d)\kappa)$, with the parameter $d \in [0,1]$, mean $\mu=\kappa$, and standard deviation $\sigma = d\kappa/\sqrt{3}$.
It was observed that the diffusion coefficient $\kappa$ had a stronger influence than the parameter $d$ on the final diversity of the system.
Furthermore, the results showed that in many configurations the optimal diffusivities were usually those corresponding to the lower cutoff introduced by a parameter value $d < 1$.
For this reason, here we focus on the uniform distribution with $d = 1$, which allows to reveal the more interesting cases when the optimal diffusivity is at some value larger than the lower cutoff $\kappa_j=0$.
In addition, we have checked that such a distribution provides results equivalent to those obtained from more realistic distributions, such as the normal distribution~\cite{HNMP-2018, HNMP-2020}.

The demographic processes are affected by the finite-range competitive interactions. 
At a given time $t$, when there are $N(t)$ bugs in the system, the generic bug $i$ [$i = 1, \dots, N(t)$] can reproduce or die according to Poisson processes with rates $r_b^i$ and $r_d^i$ (probabilities per unit of time), respectively, given by~\cite{EHG-2004}
\begin{equation}
  \begin{split}
    r^{i}_{b} &= \max\left( 0, r_{b0} - \alpha N_{R}^{i} \right) \, ,
    \\
    r^{i}_{d} &= r_{d0} \, .
   \label{eq_rates}
  \end{split}
\end{equation}
Here, $r_{b0}$ and $r_{d0}$ represent the intrinsic birth and death rates of an isolated bug, respectively.
While the death rate of bug $i$ is assumed to be constant and equal to that of an isolated bug, the reproduction rate $r^{i}_{b}$ decreases proportionally to the number of neighbors $N_{R}^{i}$ present within the finite interaction range $R$ ($R \ll L$) from bug $i$;  the parameter $\alpha > 0$ measures the intensity of the competition for resources.
Thus, the larger is the competition for resources, the smaller is the probability for reproduction.
The function $\max(.)$ ensures that $r^{i}_{b} \ge 0$, i.e. $r^{i}_{b} = 0$ when the number of neighbors $N_{R}^{i} > r_{b0}/\alpha$.

From Eqs.~\eqref{eq_rates} one can determine the critical number of neighbors within the interaction radius $R$ for which the reproduction and death of a bug are equally probable,
\begin{equation}
	N_{R}^{*} = \frac{r_{b0} - r_{d0}}{\alpha} \equiv \frac{\Delta_{0}}{\alpha} \, .
	\label{eq_NR}
\end{equation}
The quantity $\Delta_{0} = r_{b0} - r_{d0}$ is the intrinsic Malthus net growth rate of the system.
For the $i$th bug, when $N_{R}^{i}<N_{R}^{*}$, reproduction is more probable than death, while for $N_{R}^{i}\ge N_{R}^{*}$ death becomes more likely.

Notice that in the case of reproduction, the newborn is placed at the same location of the ancestor, which leads to reproductive spatial correlations~\cite{Young-2001}.
Furthermore, the newborn inherits also the ancestor's diffusion coefficient.

The time evolution of the system is simulated using the Gillespie algorithm.
At each time step, the most probable demographic event (birth or death for a certain bug) is selected.
After the demographic event has been realized, each bug $j$ makes a random step extracted from a suitable Gaussian distribution, resulting on a long time scale in a Brownian diffusion process characterized by the diffusion constant $\kappa_{j}$~\cite{HHL-2012}.
For a detailed description of the numerical algorithm see Refs.~\cite{HHL-2012,EHG-2015}.

From the studies of the homogeneous model, where all the bugs have the same diffusion coefficient $\kappa$, it is known that the interplay between reproductive correlations, limited ability of the offsprings to move away from their ancestor, and most importantly non-local competitive interactions between individuals, can under certain conditions lead to a periodic clumped spatial distribution of the organisms.
The appearance of the periodic clustering is well described with the \MF approach~\cite{CL-2004, EHG-2004,EHG-2015}. 
Using the linear stability analysis one can obtain the following condition for the periodic pattern formation,
\begin{equation} \label{pattern}
\frac{ 2 R^2 \Delta_0 }{ \kappa } > \nu_c \, ,
\end{equation}
with $\nu_c \approx 168.4$ in a one-dimensional and $\nu_c \approx 370.384$ in a two-dimensional system.
The condition~(\ref{pattern}) reveals that for fixed values of $R$ and $\Delta_0$ the diffusion coefficient $\kappa$ has to satisfy the condition $\kappa < \kappa_c$, where the critical diffusion coefficient $\kappa_c$ is given by
\begin{equation} \label{kappaC}
     \kappa_{\mathrm{c}} = \frac{ 2 R^2 \Delta_0 }{ \nu_{\mathrm{c}} } \, .
\end{equation}
{The finite range character of the competition process is a relevant feature of the model that can lead to pattern formation also in the presence of advection~\cite{EHG-2004, DeOliveira-2021a}.
In general, any finite-range interaction in some ongoing (e.g. competition, reproduction, or diffusion) process can produce relevant effects on the stationary state, even if the most relevant one remains that associated with finite-range competition~\cite{Piva-2021a}.}

In a heterogeneous system, each diffusion coefficient $\kappa_j$ satisfying the condition \eqref{pattern}, can lead to cluster formation;
typically inside a cluster all the bugs have the same diffusion coefficient.
In principle, each cluster may be characterized by a different diffusion coefficient~\cite{HNMP-2020,HNMP-2018}.

While in the previous works the initial density of bugs was assumed to be constant across the domain, here we study a system where the dispersal-structured population of bugs is initially concentrated in a small clump of radius $r_{0}$ in the middle of the otherwise empty spatial domain.
Inside the initial cluster, the density of bugs is assumed to be constant.
At least a fraction of the individuals in the population have undercritical diffusivities $\kappa_{j} < \kappac$, characteristic to a spatially periodic pattern.
In the following, the number of different diffusivities present in the system will be referred to as the diversity $D$ (the initial diversity coincides with the initial number of bugs, $D(t=0) \equiv N_0$).

We investigate how the colonization process of an initially empty area of finite size affects the selection process, i.e. the combined effects of the spreading process~\cite{cross_hohenberg-1993,Ganan-kessler-2018,Saarloos-2003,Malley_et_al_2009} and the boundary conditions.

To this aim, we consider different geometrical configurations and boundary conditions: 
a circular domain with radius $r$ and reflecting boundary conditions,
a square domain of area $L \times L$ and a one-dimensional system of linear size $L$, employing periodic boundary conditions in the last two cases. 
For any configuration, we assume that the initial spot is much smaller than the simulation domain.

As we will discuss below in detail, the numerical simulations carried out for the different geometrical configurations highlight robust features that depend on the relevant spatial scales of the system, namely the initial spot size $r_0$, the domain size, and the interaction range $R$.
Unless indicated differently, in the numerical simulations we fix the interaction radius to $R=0.1$ and the demographic parameters to $r_{b0} = 1.0$, $r_{d0} = 0.1$, and $\Delta_{0} \equiv r_{b0} - r_{d0} = 0.9$, in order to minimize the fluctuations in the population size -- see also Ref.~\cite{HHL-2012}.
Using the values of $\nu_{\mathrm{c}}$, $\Delta_0$, and $R$ given above, one finds $\kappa_{\mathrm{c}} \approx 1.07\times 10^{-4}$ in one-dimensional and $\kappa_{\mathrm{c}} \approx 4.86\times 10^{-5}$ in two-dimensional systems.
Notice that a small value of $\Delta_0$, corresponding to relatively large value of the death rate $r_{d0}$, may drive the system to extinction due to the large demographic fluctuations, even when the condition in Eq.~(\ref{pattern}) is fulfilled.

\section{Results}
\label{Sec_num_exps}

\subsection{Two-dimensional circular domain with reflecting boundary conditions}
\label{sec:circular}

\begin{figure*}[t!]
	\centering
	\includegraphics[width=17.5cm]{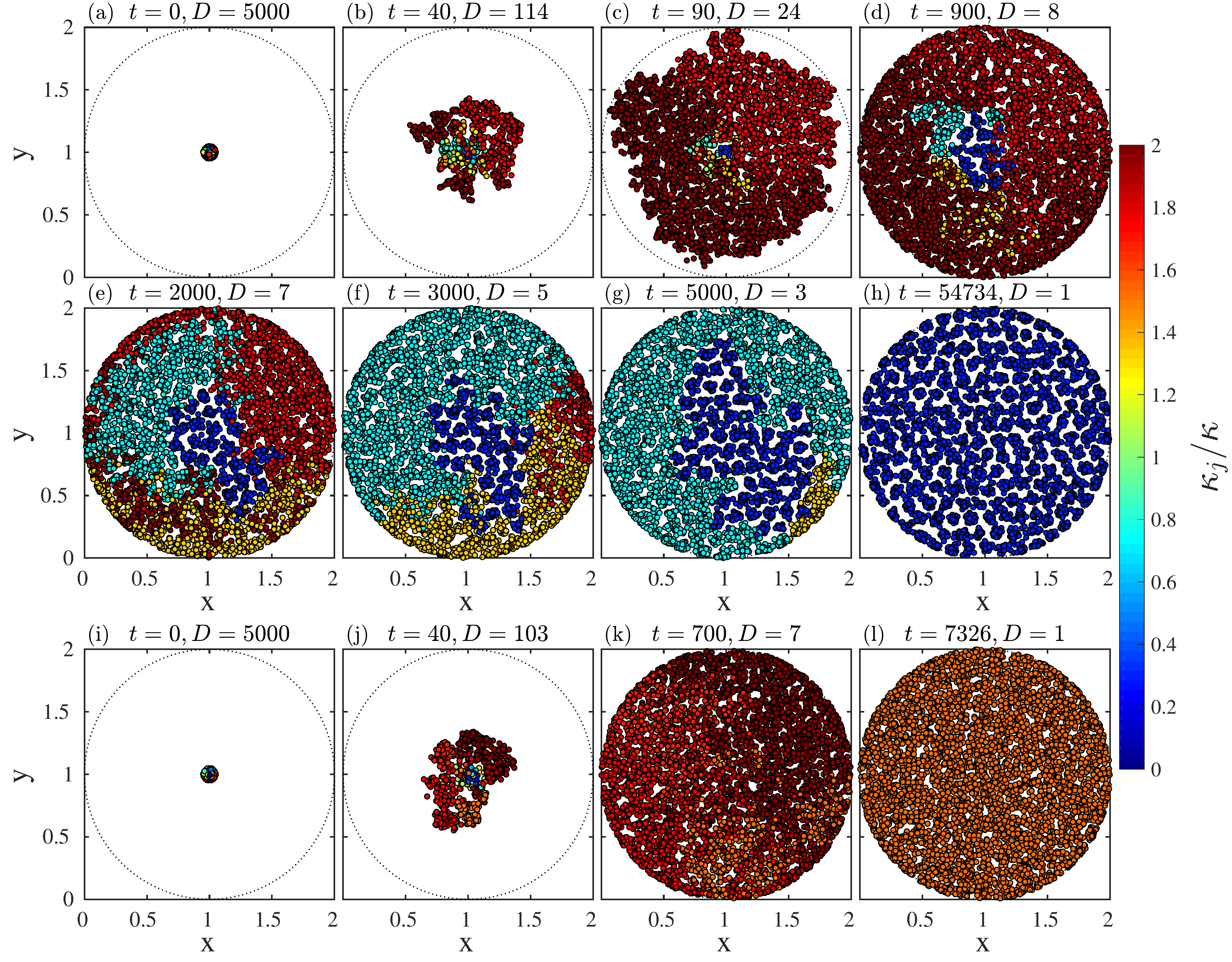}
	\caption{
    	Snapshots at various times $t$, whith the correspoding diversities $D$, from two realizations of the spatio-temporal evolution of a dispersal-structured population of bugs with initial average diffusivity $\kappa = 10^{-4}$ across a two-dimensional circular domain of radius $r = 10 \, R$ with reflecting boundary conditions.
	    The populations of bugs is initially concentrated in a central spot of diameter $2r_0=R=0.1$.
	    Panels (a)-(h): 
	    waves of different species expand across the system from the initial spot:
	    first faster ones diffuse into the empty region, then slower species gradually replace the faster ones, until the slower species with $\kappa_j = 3.05\times 10^{-5}$ wins the competition process.
	    Panels (i)-(l): 
	    the dominance of fast bugs gained after the colonization process has been maintained, as a few fast bugs with high diffusivity survive for a long time, panel (k), until eventually only the fast species with $\kappa_j = 1.5 \times 10^{-4}$ survives -- since $\kappa_j$ is overcritical no periodic spatial structure is formed.
    	Colors of bugs encode different diffusivities according to the legend.
    } 
	\label{fig_01}
\end{figure*}

Let us first consider a two-dimensional circular domain of radius $r$ with reflecting boundary conditions.
We use this configuration for illustrating the main features observed in the selection process accompanying the initial colonization process, the following consecutive front invasions, and the relaxation to the stationary state.
We assume that initially there are $N_0=5000$ bugs in the system, located in the center of the domain in a spot of radius $r_0$, characterized by a uniform diffusivity distribution $P(\kappa_j)$ in the range $\kappa_j \in [0, 2\kappa]$, with $\langle\kappa_{j}\rangle = \kappa = 10^{-4}$.
The chosen diffusivity range contains both individuals with undercritical diffusivities $\kappa_{j} < \kappac = 4.86 \times 10^{-5}$, which can develop a spatially periodic density pattern, and individuals with overcritical diffusivities $\kappa_{j}>\kappac$.
Figure~\ref{fig_01} shows two realizations of the spatio-temporal evolution of the system, leading to different final scenarios.
In both cases the simulation starts from a population of $N_0$ bugs located randomly within a spot of diameter $2 r_0 = R = 0.1$ in the middle of a circular domain of radius $r = 10 \, R$.

When $N_{0} \gg N_{R}^{*}$, the initial phase of the time evolution is always characterized by a fast decrease of the population and the diversity $D$ decreases correspondingly as a consequence of the demographic decline.
At the same time, faster bugs start a colonization process: the larger the diffusivity of the bugs, the larger their advantage in colonizing the empty part of the domain outside the initial spot (the invasion process is discussed in greater detail in Sec.~\ref{sec_Gk_vel}).
During this phase, there is a further decrease of diversity due to the continuing competition, taking place mostly in the center of the domain, accompanied by the demographic dominance of some faster types of bugs that keep expanding into the empty part of the domain.
The growth of the population of the faster bugs stops when the boundaries of the domain are reached.
While this first phase is common in all realizations, the following time evolution can differ substantially in each realization, due to the stochastic nature of the system dynamics.

In the first realization, shown in Fig.~\ref{fig_01}, panels (a)-(h), a slower type of bugs forming a periodic pattern (with $\kappa_j < \kappa_c$) wins the competition. 
In the second example, shown in Fig.~\ref{fig_01}, panels (i)-(l), a faster type of bugs with $\kappa_j > \kappa_c$ eventually remains as the only species in the system.
In order to gain a more complete overall picture of the underlying selection processes and the possible different final configurations, it is useful to study the probability distribution of diffusivities $P(\kappa_{j})$.

\begin{figure*}[t!]
\centering
\includegraphics[width=16.5cm]{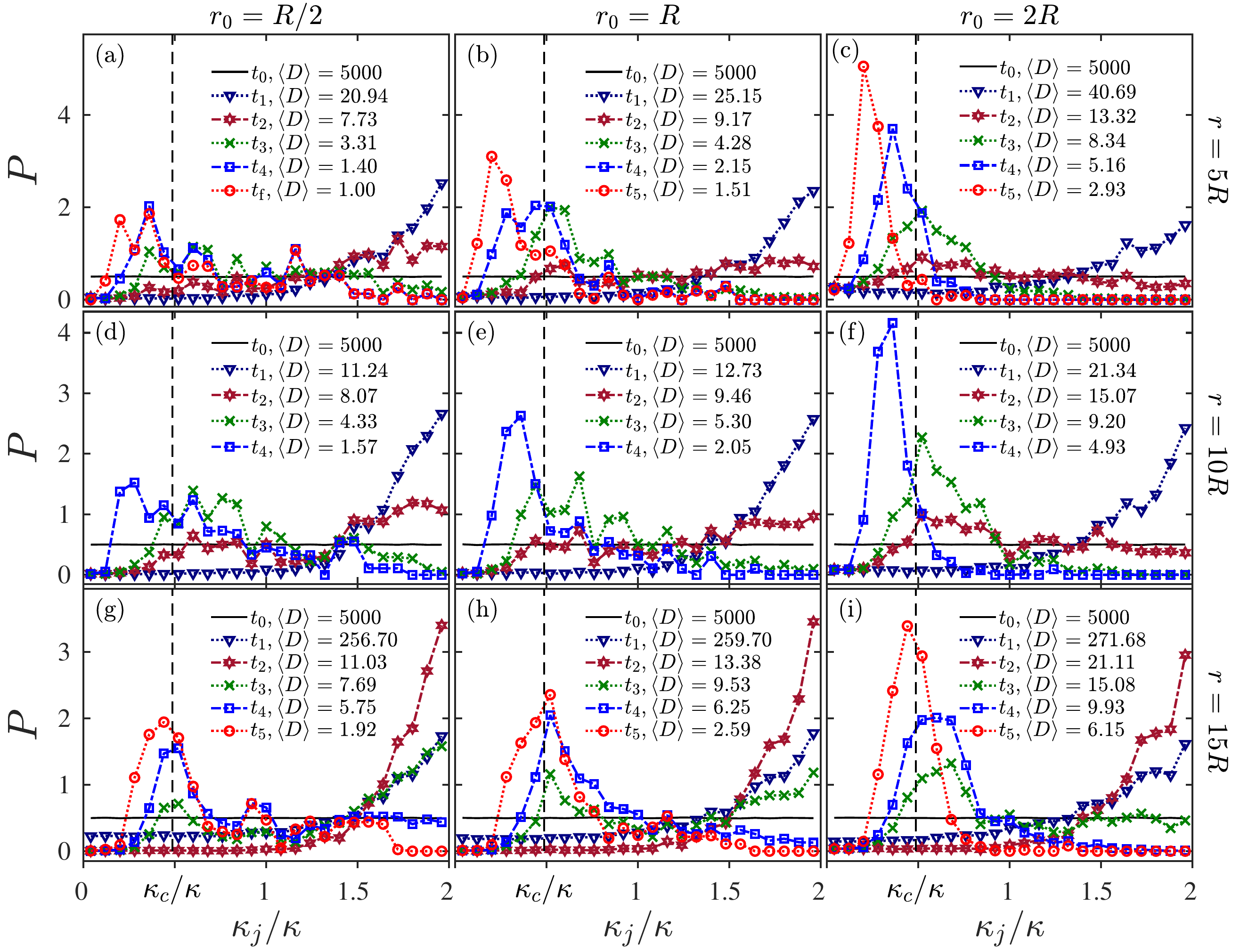}
\caption{
	Diffusivity distribution $P(\kappa_{j})$ at different times $t$ for a dispersal-structured population with initial average diffusivity $\kappa=10^{-4}$ spreading across a two-dimensional circular domain, for different domain radii $r$ and initial spot radii $r_{0}$ indicated on the right and top side, respectively. 
    The average diversity $\langle D \rangle$ at the various times is also given.
	Curves are obtained averaging over 100 realizations.
    The critical diffusivity value $\kappac = 4.86 \times 10^{-5}$ is marked with a dashed lines.
} 
\label{fig_02}
\end{figure*}

The probability distributions $P(\kappa_{j})$ at different times, obtained by averaging over 100 realizations, are shown in Fig.~\ref{fig_02}, for different values of the initial spot radius $r_0 = R/2, R, 2R$ and domain radius $r = 5R, 10R, 15R$ (larger domain sizes produce results equivalent to those for $r = 15R$).

Let us first look at the distributions $P(\kappa_j)$ in Fig.~\ref{fig_02}-(d), corresponding to the configuration used for the realizations in Fig.~\ref{fig_01}.
Following the time evolution, one can notice that starting from the uniform distribution $P(\kappa_j)$, a maximum appears at high diffusivities: the larger the diffusivity, the higher is $P(\kappa_j)$.
This reflects the colonization process carried out by the fast bugs, observed in the realizations shown in Fig.~\ref{fig_01}.
As time goes on, the maximum of $P(\kappa_j)$ at larger diffusivities starts to disappear, while a new maximum appears at smaller values around the critical diffusivity $\kappac$.

While the faster bugs are more effective in colonizing the empty area, the slower bugs, if they manage to spread and start a new cluster in an already populated area, have a competitive advantage.
In general, we find that the slower the bugs, the stronger the clusters they form; however, at the same time, they are less effective in the invasion, confirming results previously obtained in Ref.~\cite{HHL-2013}. 
This is the reason why the maximum of the distribution $P(\kappa_j)$ appears at intermediate values, around the critical value $\kappa_c$, which represents a good compromise.
The outcome of the first realization shown in Fig.~\ref{fig_01}, panels (a)-(h), where the final diffusivity is $\kappa_{j}=3.05\times 10^{-5}$, 
well matches the maximum at intermediate $\kappa$ shown in Fig.~\ref{fig_02}-(d).

However, due to statistical fluctuations, bugs with diffusivities around the critical value $\kappac$, which should give them a competitive advantage, can also disappear at some point of the time evolution and only faster bugs with $\kappa_j \approx 1.4 \times 10^{-4}$ survive --- see the corresponding peak in the diffusivity distribution $ P(\kappa_j$) in Fig.~\ref{fig_02}-(d).
The second realization shown in Fig.~\ref{fig_01}, panels (i)-(l), in which the final surviving species has $\kappa_j = 1.5 \times 10^{-4}$, is an example of such a situation.
Figure~\ref{fig_01}-(j) shows some clusters of slower bugs appearing for some time and disappearing afterwards because of statistical fluctuations.
Then the remaining bugs have high diffusivities, see Fig.~\ref{fig_01}-(k) at $t=700$, from which eventually only one survives.

The other panels in Fig.~\ref{fig_02}, corresponding to different initial spot and system size, exhibit a time evolution similar to that of panel (d) discussed above.
At the same time, despite the similar dynamics, inspection of the final diffusivity distributions shows some relevant differences and dependencies on the system length scales:
increasing the initial spot size $r_0$ increases in turn the probability that a bug with a diffusivity $\kappa \lesssim \kappac$ survives, while the maximum of the distribution at the highest diffusivity gradually decreases;
instead, as the system size $r$ is decreased, the maximum of the final diffusivity distribution at $\kappa \lesssim \kappac$ moves towards smaller values of diffusivity.
Thus in both cases, increasing $r_0$ or decreasing $r$, smaller diffusivities acquire an advantage in the natural selection process.

Comparison among the diffusivity distributions $P(\kappa_j$) shown in Fig.~\ref{fig_02} suggests that the ratio $r_0/r$ represents a main length-related parameter of the system, describing the combined effect of $r_0$ and $r$\,:
panel (c), associated with the largest ratio considered, $r_0/r = 2/5$, and panel (g), associated with the smallest ratio $r_0/r = 1/30$, correspond the two distributions most different from each other among all those in Fig.~\ref{fig_02} (for the role of the parameter $r/r_0$ see also Fig.~\ref{fig_07} in Sec.~\ref{sec:1D}).

\begin{figure*}[t!]
\centering
\includegraphics[width=16.5cm]{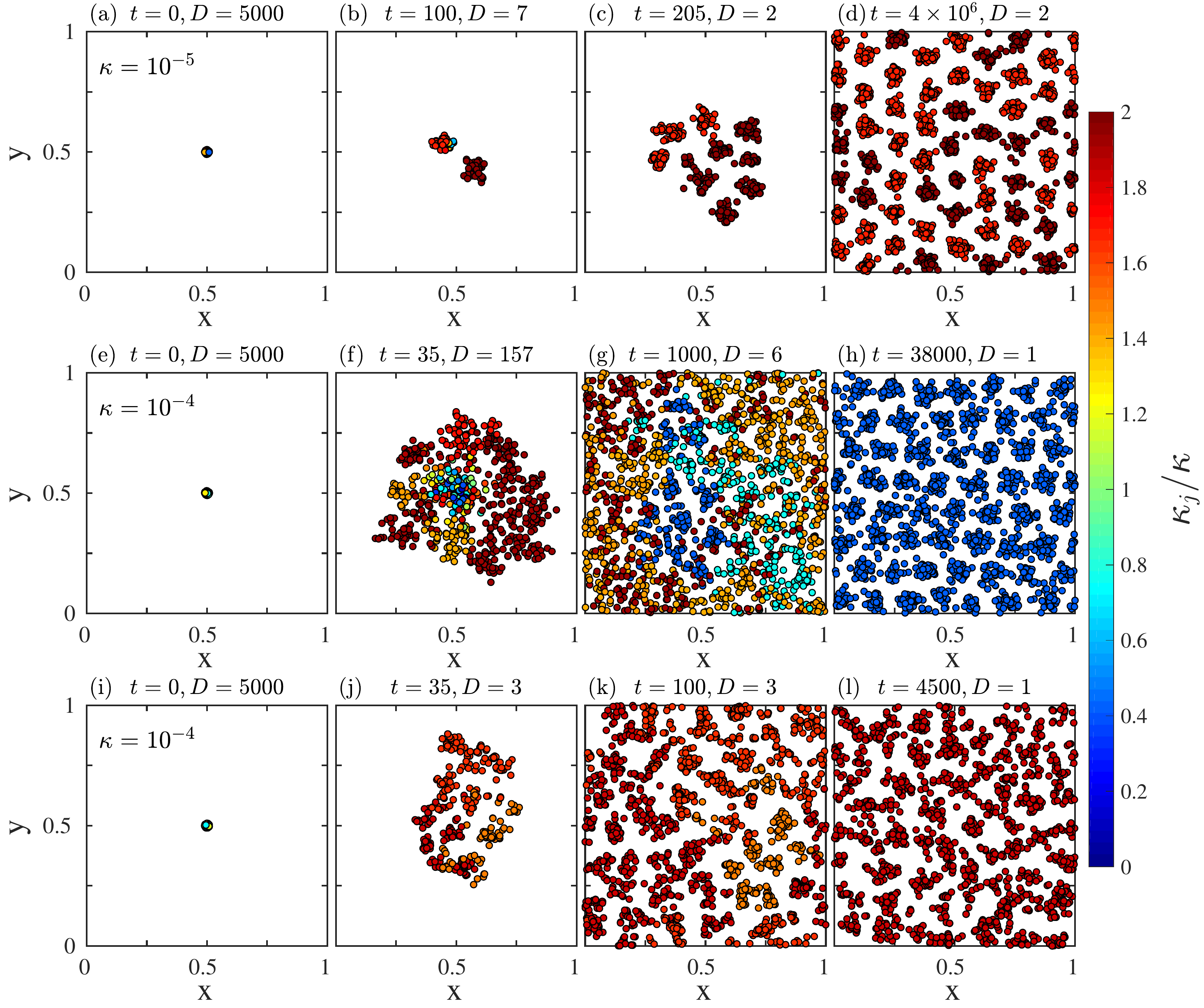}
\caption{
    Snapshots at various times $t$, whith the correspoding diversities $D$, from the space-time evolution of a dispersal-structured population of bugs initially localized within an spot of width $2r_{0}=R/5$ and diffusing across a square two-dimensional domain of linear size $L=1$ with periodic boundary conditions.
	Panels (a)-(d): realization with undercritical initial mean diffusivity $\kappa=10^{-5}$ leading in the long time limit to a periodic arrangement of bugs with two diffusivities $\kappa_{j} = 1.67 \times 10^{-5}$ and $\kappa_{j} = 1.96 \times 10^{-5}$. 
	Panels (e)-(l): realization with overcritical initial mean diffusivity $\kappa=10^{-4}$, leading to a spatially periodic pattern at $t=38000$ containing the species with undercritical diffusivity $\kappa_j=4.2\times 10^{-5}$. 
	Panels (i)-(l): another realization with overcritical initial mean diffusivity $\kappa=10^{-4}$ that has a final state at $t=4500$ with a random distribution of bugs characterized by the overcritical diffusion coefficient $\kappa_{j}=1.83 \times 10^{-4}$. 
	Colors of bugs encode different diffusivities according to the legend.
} 
\label{fig_03}
\end{figure*}

\subsection{Square domain with periodic boundary conditions}
\label{Sec:square}

Here we study a square domain configuration with periodic boundary conditions.
The comparison with the results of the previous section allows to check the possible role of boundary effects.
We consider a small system with linear size $L=1$ and a population initially composed of the same number of bugs $N_{0} = 5000$ localized within a spot of diameter $2r_{0}=R/5$.
We study the case of a population with mean diffusivity $\kappa=10^{-4}$, where both sub- and supercritical diffusivities are present, as in the examples studied in the previous section, and that of a mean diffusivity $\kappa=10^{-5}$, containing only subcritical diffusivities $\kappa_{j} < \kappa_{\mathrm{c}} = 4.86 \times 10^{-5}$. 

\begin{figure*}[t!]
	\centering
	\includegraphics[width=7.8cm]{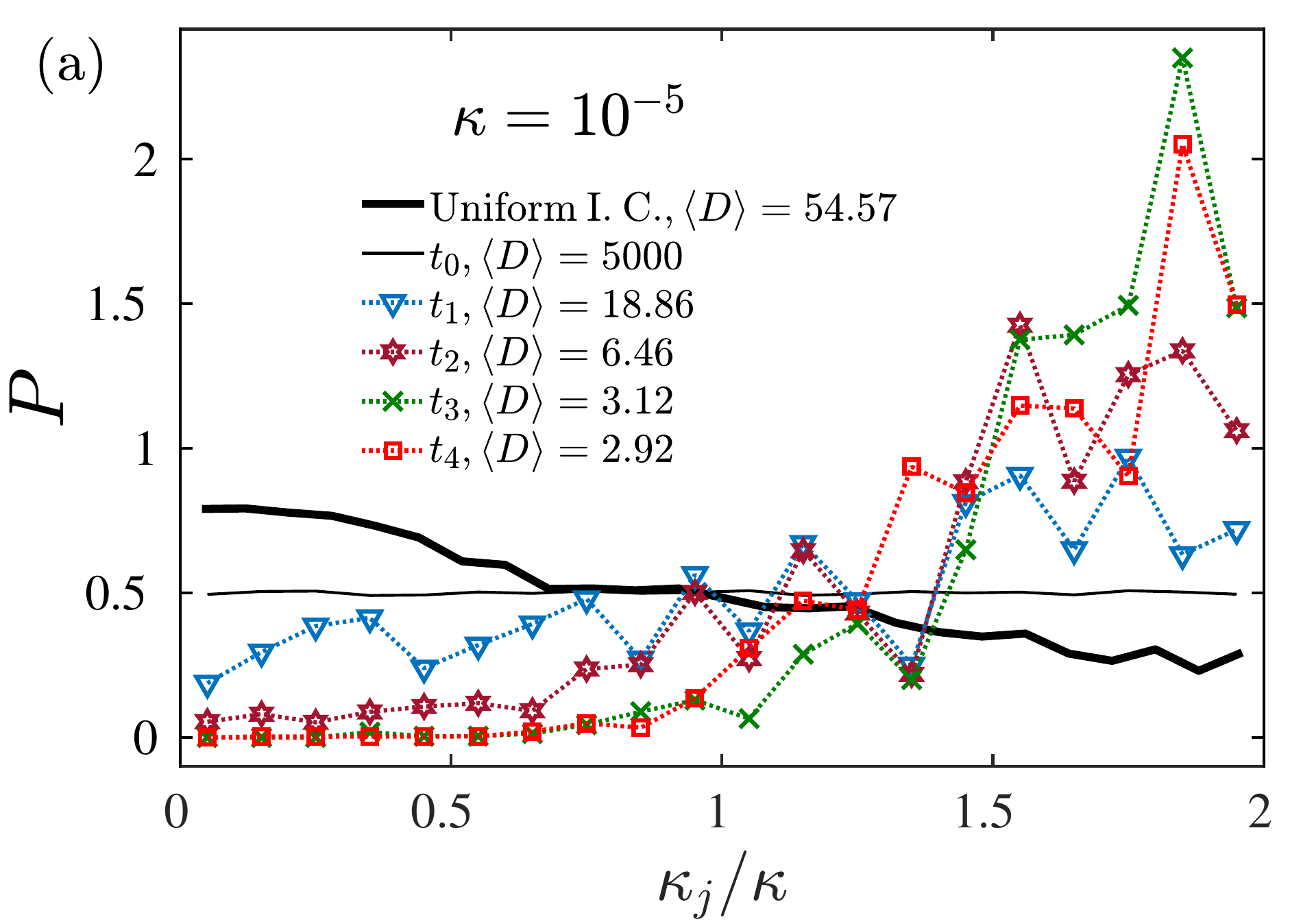}\hspace{0.1cm}
	\includegraphics[width=7.8cm]{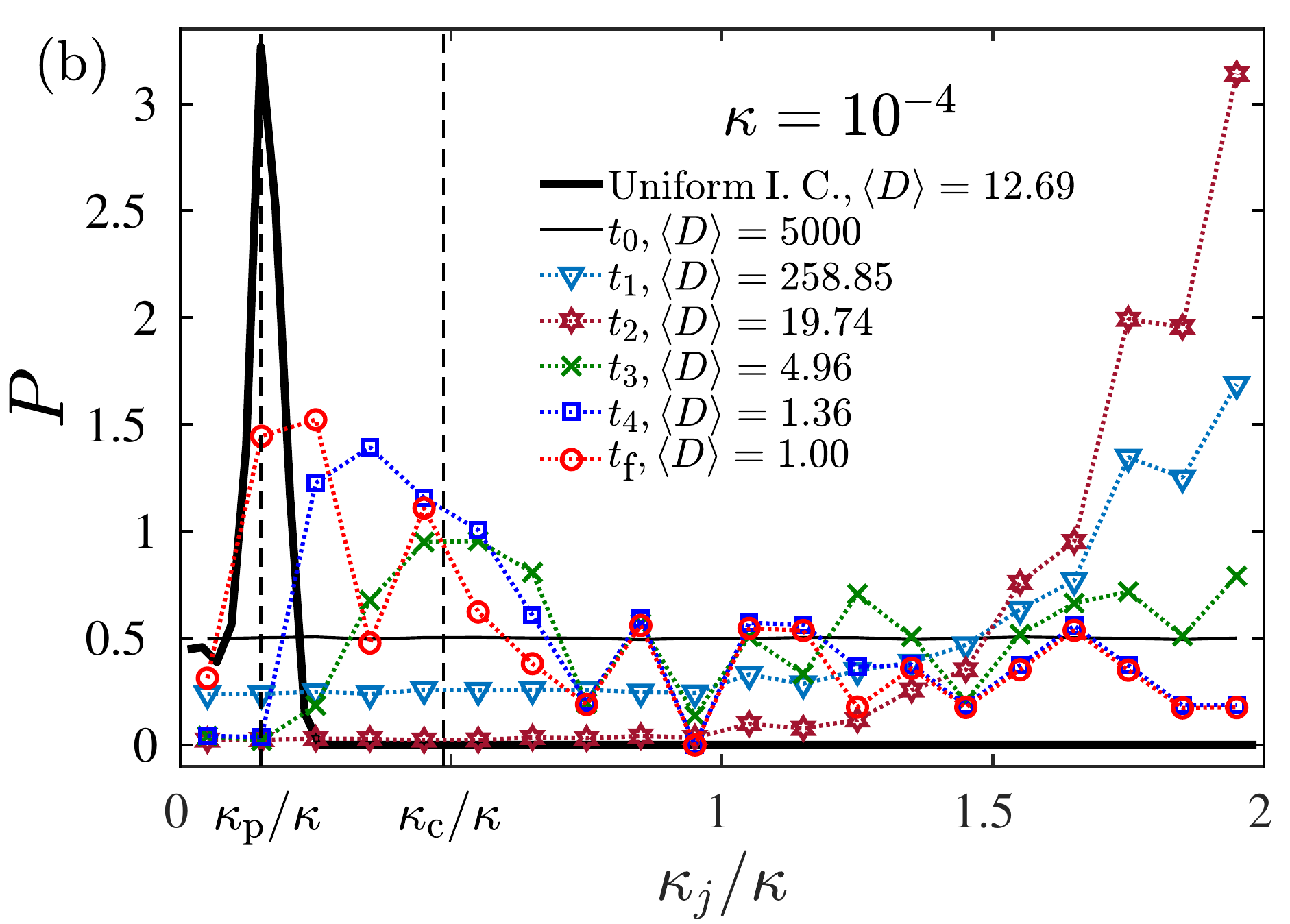}	\\
	\caption{
	  Time evolution of the diffusivity distribution $P(\kappa_{j})$ of a heterogeneous population with an initial number of bugs $N_{0} = 5000$.
	  Panel (a): initial mean diffusivity $\kappa = 10^{-5}$.
	  Panel (b): initial mean diffusivity $\kappa = 10^{-4}$.
	  Bugs are initially located within a central spot of width $2r_{0} = R/5$ and diffuse across a squared domain of area $L \times L \> (L=1) $ with periodic boundary conditions. 
	  The distributions are compared with the long-time limit of the probability distributions of a population starting from a uniform spatial density of bugs (solid black lines, reduced by a factor 3.5 in panel (b)) with $\langle D \rangle = 54.7$ at $t \approx 6 \times 10^6$ in panel (a) and $\langle D \rangle = 12.69$ at $t \approx 9 \times 10^{6}$ in panel (b). 
	  Curves are obtained averaging over 100 realizations, solid black curves are obtained as an average over 150 realizations~\cite{HNMP-2018,HNMP-2020}.} 
	\label{fig_04}
\end{figure*}

Let us start with the case $\kappa=10^{-4}$. Also for this study we present two examples of realizations:
one where a species of slower bugs with a subcritical diffusivity wins the competition process, Fig.~\ref{fig_03}, panels (e)-(h); 
and one where a species of faster bugs with an overcritical diffusivity survives, Fig.~\ref{fig_03}, panels (i)-(l).
Analogously to the simulations discussed in Sec.~\ref{sec:circular}, in Fig.~\ref{fig_03}, panels (e)-(h), we observe consecutive waves of bugs:
the first waves associated with larger diffusivities correspond to the colonization of the empty area, while the following waves, associated with smaller diffusivities, correspond to invasions of other species across an already populated region.
Eventually, a slow species wins, forming a regular spatial pattern across the domain.
Instead, in the realization shown in Fig.~\ref{fig_03}, panels (i)-(l), the initial dominance of the faster bugs gained after the colonization is maintained in the long time limit, because slower bugs disappear from the system due to statistical fluctuations.
Notice that the surviving diffusivity is larger than the critical one, $\kappa_{j}=1.83 \times 10^{-4} > \kappac$, so the spatial distribution of bugs is not periodic; however, one can notice some clustering and spatial structures due to spatial correlations induced by finite-range competition and reproductive correlations.

Figure~\ref{fig_04}-(b) shows the corresponding diffusivity distributions $P(\kappa_{j})$ for $\kappa=10^{-4}$, obtained by averaging over 100 realizations.
The time evolution of the distributions $P(\kappa_{j})$ confirms an initial enhancement (from $t=0$ to $t \approx t_{2}=100$) of the distribution tail, $\kappa_{j}/\kappa \gtrsim 1.4$, due to faster bugs colonizing the empty area, where their reproduction is favored. 
At time $t \approx t_{3} = 2000$, the density has developed a clear maximum at $\kappa \approx \kappac$, signalling the spreading and proliferation of bugs with diffusivities around the critical diffusivity. 

Let us now compare the results for the square domain with the case of a circular domain with reflecting boundary conditions studied above.
The comparison is more appropriate if done with Fig.~\ref{fig_02}-(a), since it has a comparable domain size $2r=1$ and a smaller initial spot size.
As one can see, the time evolutions and the final steady shapes of the distributions $P(\kappa_j)$ show no significant differences.
This demonstrates the robustness of the results obtained with respect to the type of boundary conditions.

The natural selection process for an initial mean diffusivity $\kappa = 10^{-5}$ presents substantial differences with respect to the case of the higher diffusivity $\kappa = 10^{-4}$, compare the realization depicted in Fig.~\ref{fig_03}, panels (a)-(d), with the realizations discussed above in panels (e)-(h) and (i)-(l). 
Because all the bugs have low diffusivities, it takes a long time for any of them to escape from the initial overcrowded spot and start the formation of a new cluster, i.e. there is no wave of fast bugs.
Under these conditions, intense competition and selection processes already take place inside the initial spot, depicted in Fig.~\ref{fig_03}-(a).
In fact, at time $t=100$, Fig.~\ref{fig_03}-(b), diversity has already reduced to $D=7$, while only two clusters are present in the system:
a first cluster that has evolved directly from the initial spot, populated by bugs diversified by just a few diffusivities;
and another cluster containing only one species, started by a faster bug (compare the color scale) that managed to escape from the initial spot and diffuse across the domain.
(The difference with the situation for the larger $\kappa=10^{-4}$ is evident when comparing panel (f) at the smaller time $t = 35$.)
At time $t=205$, Fig.~\ref{fig_03}-(c), the diversity has already dropped to $D = 2$.
Due to the small diffusivities, the probability that a bug manages to cross the inter-cluster region is very small~\cite{HNMP-2020,HHL-2013,HNMP-2018}. 
Thus, the two surviving species continue to colonize the domain, Fig.~\ref{fig_03}-(d), coexisting for a long time (compare the time with that of panel (h) for $\kappa = 10^{-4})$.
In this case, the time evolution of the probability distribution $P(\kappa_{j})$, shown in Fig.~\ref{fig_04}-(a), reveals that only relatively large diffusivities $\kappa_{j}/\kappa \gtrsim 0.7$ manage to survive the competition process.

The size effects of the initial spot and of the domain are best seen by comparing the final diffusivity distributions with those obtained starting from an initial uniform spatial density of bugs, represented as solid curves in Fig.~\ref{fig_04}-(a) and \ref{fig_04}-(b).
First, for $\kappa = 10^{-4}$, the range of diffusivities around the value $\kappa = \kappap$ that locates the maximum of the distribution $P(\kappa_j)$ obtained from an initial uniform spatial density of bugs are similarly favored here, but the new distribution from localized initial conditions is much broader.
Furthermore, there is now a non-zero probability that diffusivities $\kappa_j > \kappac$ can win the competition process.
For the smaller initial mean diffusivity $\kappa = 10^{-5}$, even the general trend is different, in that now the larger (rather than the smaller) diffusivities present in the system are favored.

\begin{figure*}[t!]
	\centering
	\includegraphics[width=16.5cm]{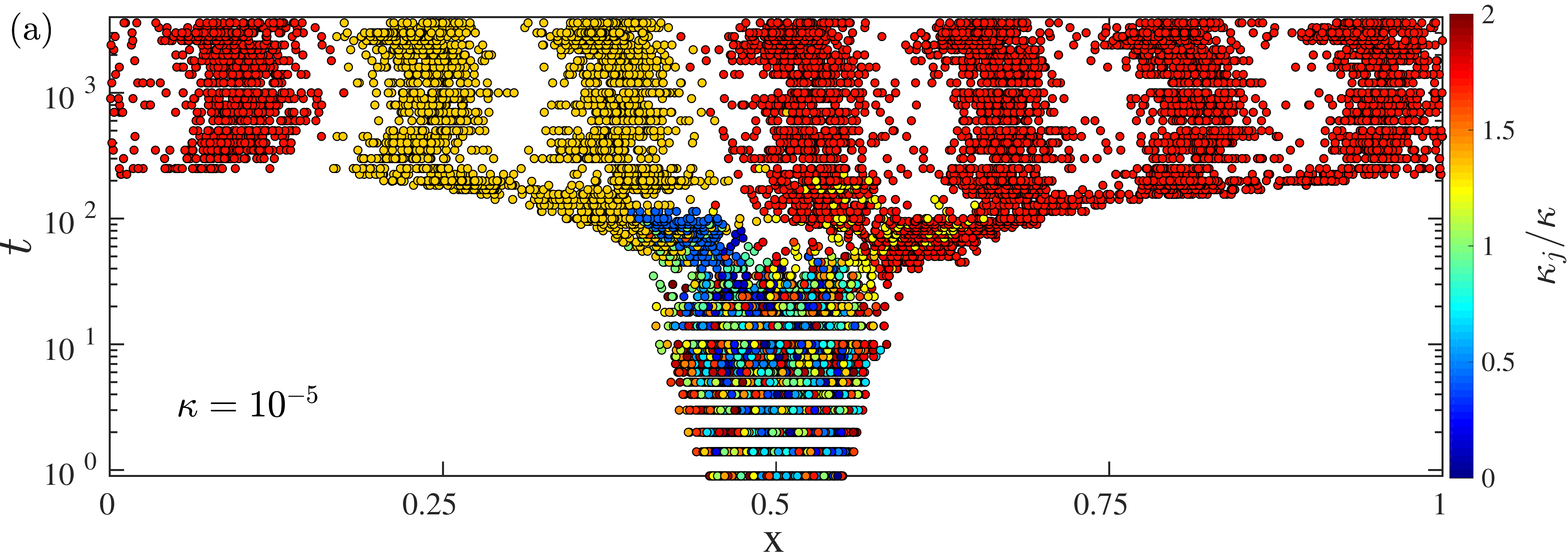}
	\includegraphics[width=16.5cm]{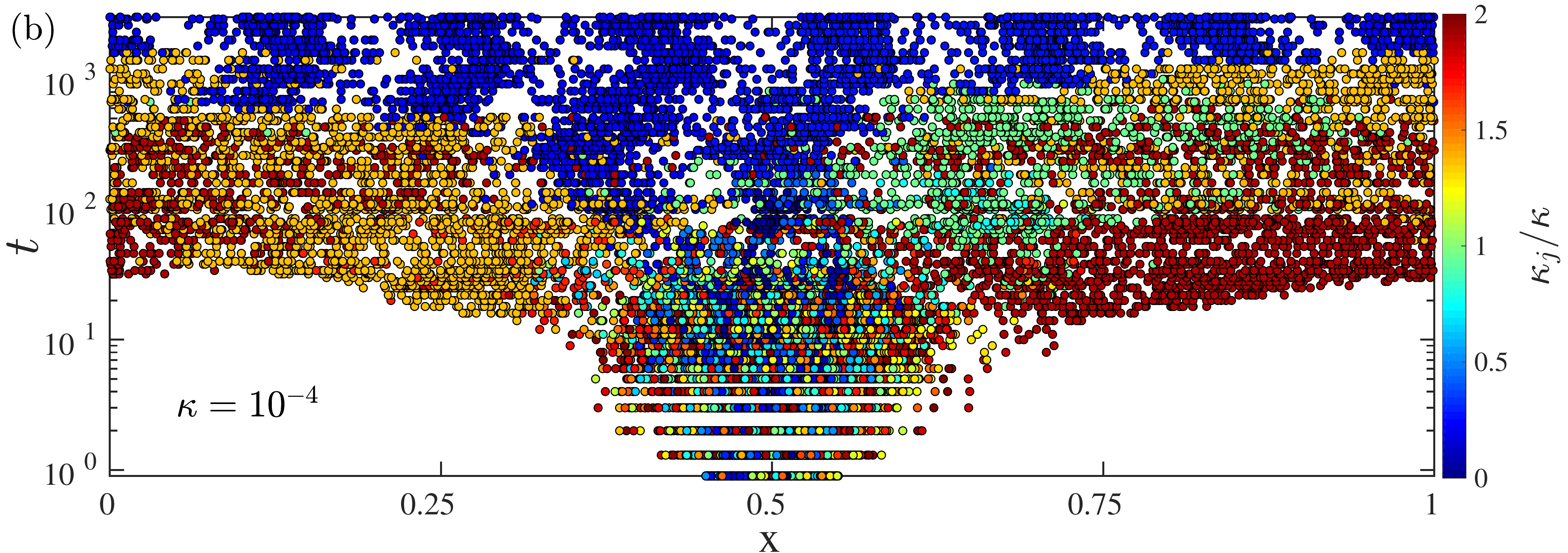}
	\caption{
	Spatio-time evolution of dispersal-structured populations of bugs, showing the colonization and invasion processes.
	The initial population contains $N_{0}=1500$ bugs placed within a spot of width $2r_{0}=R$.
	The domain has a linear size $L=1$ and periodic boundary conditions. 
	Panel (a): realization with initial undercritical mean diffusivity $\kappa=10^{-5} < \kappa_{\mathrm{c}}$. 
	By time $t \approx 3000$, the system has developed a periodic arrangement of bugs containing the two competing diffusivities $\kappa_{j} = 1.3 \times 10^{-5}$ and $\kappa_{j} = 1.7 \times 10^{-5}$. 
	Panel (b): realization with initial mean diffusivity $\kappa = 10^{-4} \approx \kappa_{\mathrm{c}}$.
	At $t \approx 3000$, the bugs distribution is characterized by a periodic pattern containing bugs with a single small diffusion coefficient $\kappa = 2.2\times 10^{-5}$. 
	Colors of bugs encode the values of the respective rescaled diffusivities $\kappa_j/\kappa$ according to the legend.
	} 
	\label{fig_05}
\end{figure*}

\begin{figure*}[t!]
	\centering
	\includegraphics[width=8.cm]{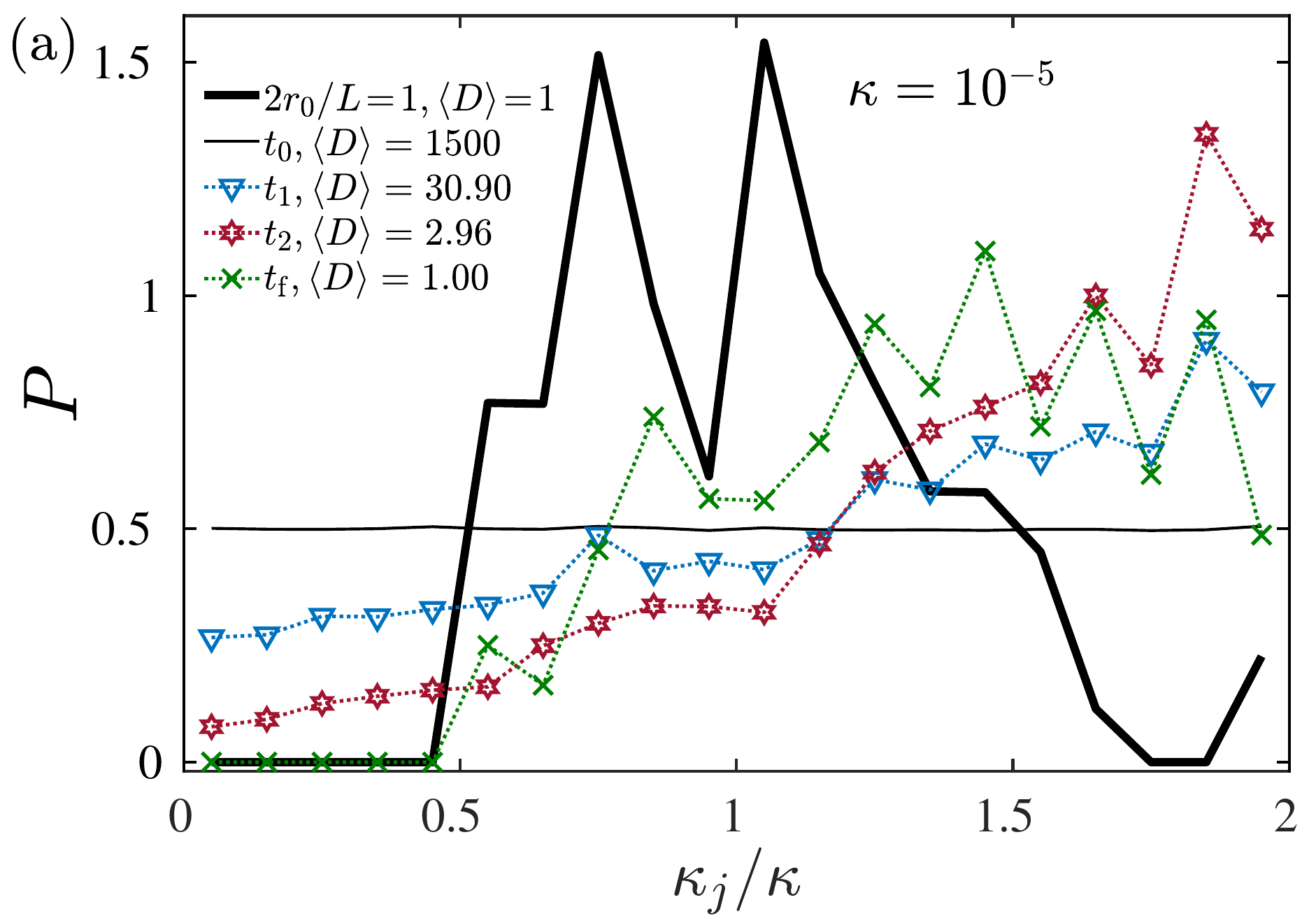} 
	\includegraphics[width=8.cm]{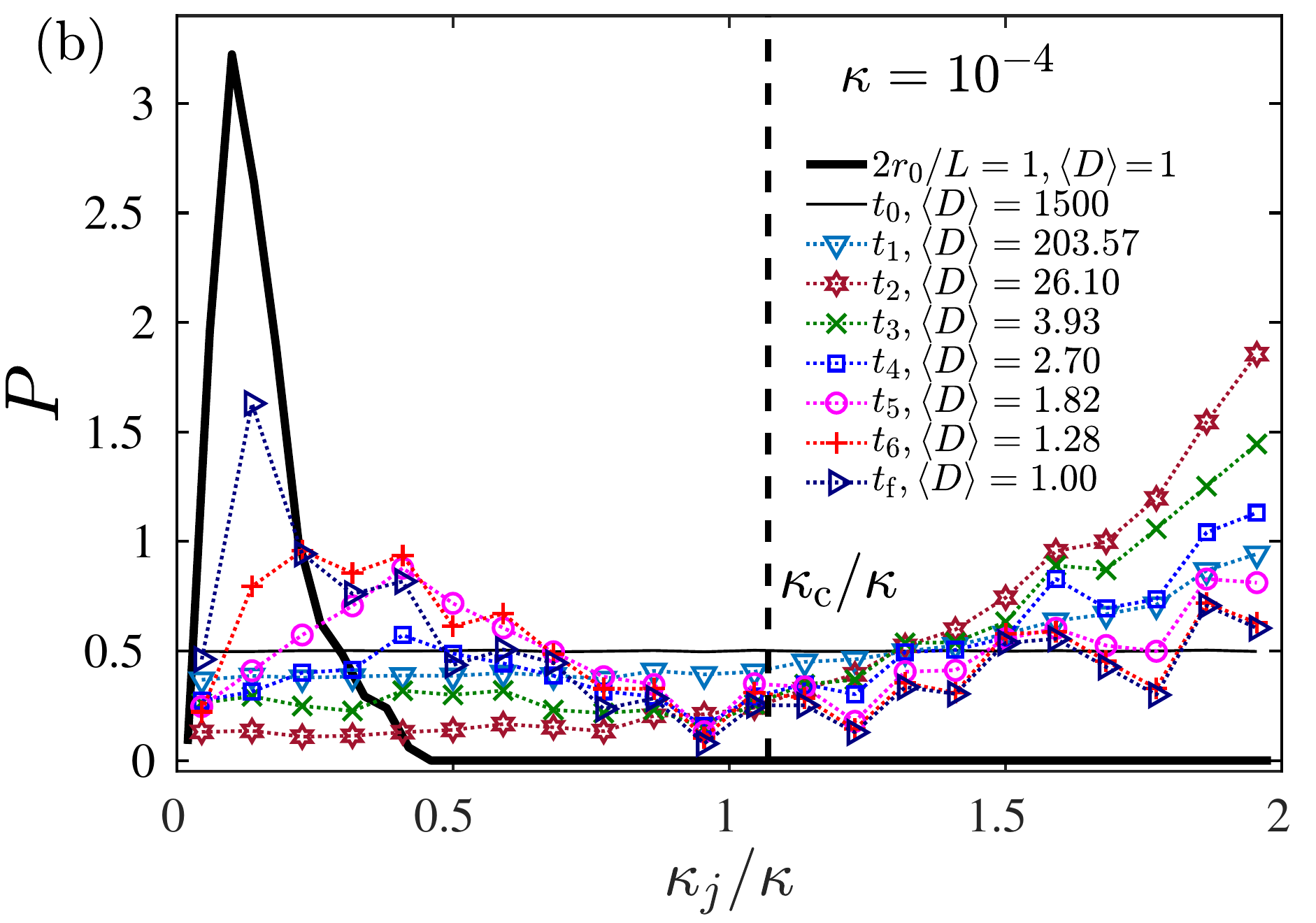}\\	
	\caption{
		Time evolution of the probability distribution $P(\kappa_{j})$ (marked colored lines, obtained by averaging over 400 runs) in the heterogeneous one-dimensional bugs model with domain size $L=1$, starting from localized spatial distributions of bugs with spot size $2r_{0}=R$.
		Panel (a): initial mean diffusivity $\kappa=10^{-5}$.
		Panel (b): initial mean diffusivity $\kappa=10^{-4}$. 
		The final diffusivity distributions obtained starting from an initial constant density of bugs ($2r_{0}/L=1$) are shown as solid black curves (reduced by a factor 2 in panel (b)). 
	} 
	\label{fig_06}
\end{figure*}


\subsection{One-dimensional domain with periodic boundary conditions}
\label{sec:1D}

In this section we consider a population of bugs in a one-dimensional domain with periodic boundary conditions,
allowing faster simulations and an interesting visualization of the space-time evolution of the system.
We start the simulation from an initial spot size of width $2r_0 = R$, containing $N_{0}=1500$ bugs, so that the initial density $N_{0}/2r_0 = 15000$ is much larger than the critical density $N_{R}^{*}/2R = 225$ (the other parameters are unchanged).

Figure~\ref{fig_05}-(a) and \ref{fig_05}-(b) show the spatio-time diagrams of two realizations for two populations of bugs with initial mean diffusion coefficient $\kappa=10^{-5}$ and $\kappa=10^{-4}$, respectively --- for comparison the critical diffusivity in a one-dimensional system is $\kappac = 1.07 \times 10^{-4}$, as pointed out in Sec.~\ref{Sec-system}.

In the simulation shown in Fig.~\ref{fig_05}-(a), corresponding to $\kappa=10^{-5}$, even the largest diffusivity present in the population is subcritical, leading to a final spatial periodic arrangement of bugs.
As observed in the previous simulations, the initial wave fronts are led by faster bugs, which can leave the initial spot more easily to spread towards the boundaries.
A periodic pattern extending through the whole domain is developed at $t\approx 300$ made up of only two species competing with each other, characterized by diffusivities $\kappa_{j}=1.7\times 10^{-5}$ and $\kappa_{j}=1.3\times 10^{-5}$.
The time needed by a heterogeneous system to relax to the stationary state with diversity $D=1$ can be very long for low values of $\kappa$; 
for the realization depicted in Fig.~\ref{fig_05}-(a), the convergence to a stationary state is achieved at $t\approx 1.4\times 10^{6}$ when the smaller diffusivity wins the competition process (not shown).

In the realization with $\kappa=10^{-4}$, shown in Fig.~\ref{fig_05}-(b), fast bugs with overcritical diffusivities $\kappa_{j} > \kappac$ develop the first wave fronts; correspondingly, the density of bugs across the domain is rather homogeneous in the time interval $t \approx (20,80)$.
A second spreading process, this time an invasion process of a domain already populated by bugs, is carried out by slower bugs with $\kappa_{j} = 2.2\times 10^{-5}$ starting at $t \gtrsim 100$.
The spreading of slower bugs is eventually successful, thanks to their advantage over faster bugs~\cite{HNMP-2018,HNMP-2020,HHL-2013}.

Notice that the dynamics of these realization closely resembles those of the two-dimensional systems studied above with the same initial mean diffusivity $\kappa$:
the realization depicted in Fig.~\ref{fig_05}-(a) for $\kappa = 10^{-5}$ is similar to that of the square domain depicted in Fig.~\ref{fig_03}, panels (a)-(d),
while the realization in Fig.~\ref{fig_05}-(b) for $\kappa = 10^{-4}$ recalls that of the circular domain with reflecting boundary conditions illustrated in Fig.~\ref{fig_01}, panels (a)-(h), or the square domain with periodic boundary conditions shown in Fig.~\ref{fig_03}, panels (e)-(h).

In Fig.~\ref{fig_06} we present the time sequences of the probability distributions $P(\kappa_j)$  corresponding to the same set of parameters used for the realizations shown in Fig.~\ref{fig_05}. 
The similarity with the time evolution of the diffusivity distributions for a square domain, depicted in Fig.~\ref{fig_04}, shows the robustness of the main features of the dynamics {and confirms the strong influence of the localized initial conditions on the shape of the diffusivity distribution}.
The only relevant difference between the one- and the two-dimensional case seems to be that for a initial mean diffusivity $\kappa=10^{-4}$ the maximum of the distribution $P(\kappa_j)$ developed at the largest available $\kappa_j = 2\kappa$ is maintained even in the final stationary distribution at $t = \tf$ in the one-dimensional case, Fig.~\ref{fig_06}-(b), but not in the two-dimensional case, Fig.~\ref{fig_04}-(b) (periodic boundary conditions) and Fig.~\ref{fig_02} (reflecting boundary conditions).

The comparison made in Fig.~\ref{fig_06} between the curves of the final distribution at $t = \tf$ and the final distributions obtained from a uniform initial spatial density of bugs (solid curves) reveals remarkable differences that demonstrate a relevant role of the size of the initial spot.
In fact, when a heterogeneous population with $\kappa=10^{-5}$ is initially distributed across the entire domain of size $L=1$, the final distribution $P(\kappa_{j})$ in panel (a) presents two peaks at intermediate values of the $\kappa_j$ range, corresponding to periodic configurations containing $7$ or $8$ clusters; the peak locations can be connected to special values of the diffusivity, see Ref.~\cite{HNMP-2020} for details. 
In contrast, the diffusivity distribution corresponding to localized initial conditions does not reveal the existence of special values of the diffusivity 
that could be associated with particular periodic arrangements of the system.

\begin{figure*}[!t]
	\centering
	\includegraphics[width=8.6cm]{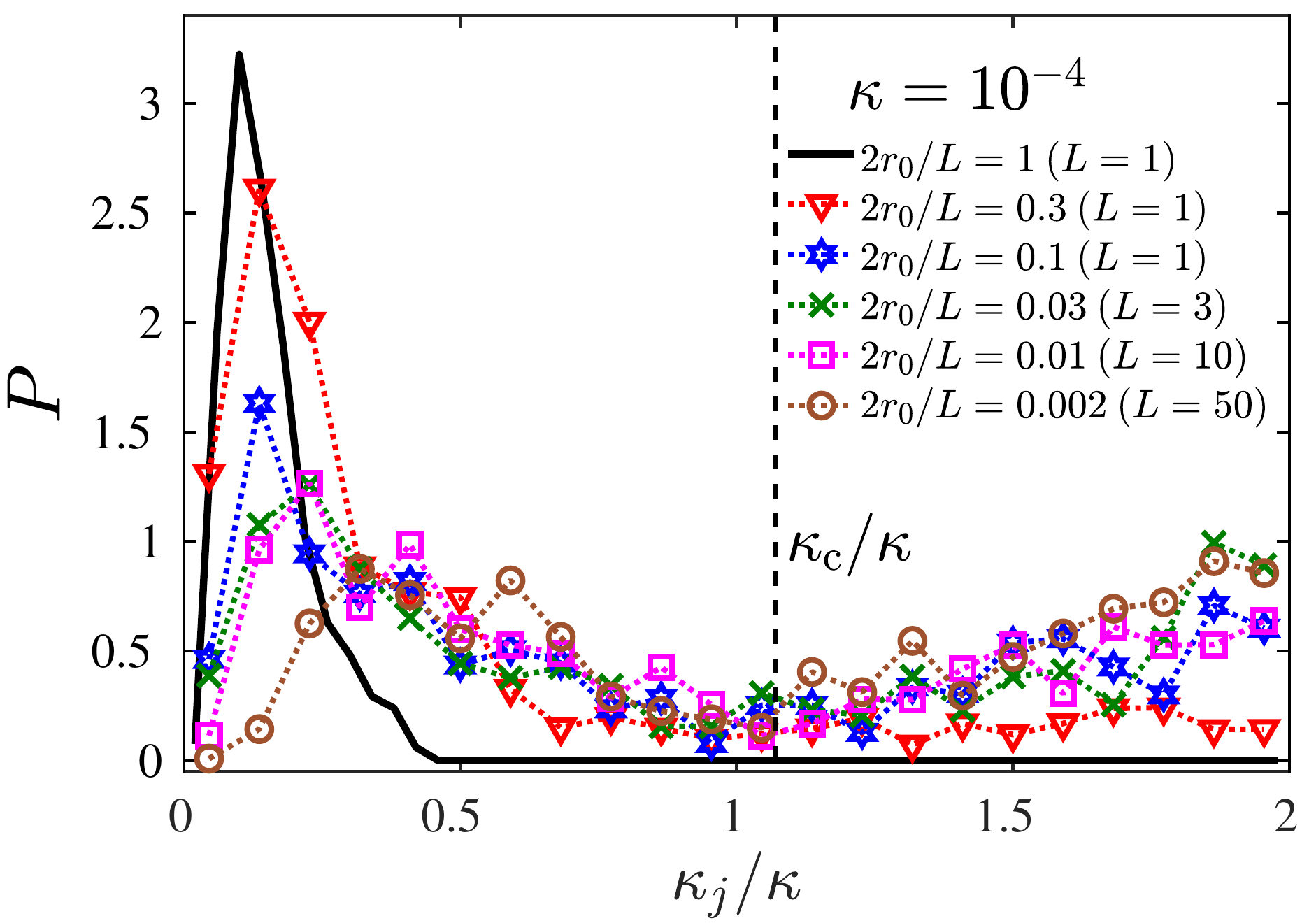}
	\caption{
	    Stationary distribution $P(\kappa_{j})$ of a one-dimensional heterogeneous systems of bugs with initial mean diffusivity $\kappa=10^{-4}$ for different values of the ratio $2r_{0}/L$ (marked lines; diversity $\langle D \rangle =2.06$ for the case $2r_{0}/L=0.002$, and $D=1$ for all the others; averaged over 400 realizations). 
	    For comparison, the figure shows also the distribution $P(\kappa_{j})$ obtained starting from an initial constant spatial distribution of bugs, $2r_{0}/L = 1$ (solid black line; distribution rescaled by a factor 2; averaged over 188 realizations~\cite{HNMP-2018,HNMP-2020}). 
        The dashed line marks the critical diffusion coefficient in a one-dimensional system. 
	} 
	\label{fig_07}
\end{figure*}

Finally, we examine the diffusivity distributions $P(\kappa_{j})$ resulting from the competition dynamics in systems with initial mean diffusivity $\kappa=10^{-4}$ for different values of the ratio $2r_{0}/L$.
We vary both the value of the domain size $L$ and initial spot width $r_{0}$.
Results are shown in Fig.~\ref{fig_07} for different values of the ratio $2r_{0}/L \in (0.002, 1)$.
As the ratio $2r_{0}/L$ is increased, the competitive advantage is given to smaller diffusivities at the expense of larger diffusivities, until already at the value $2r_{0}/L = 0.3$ the diffusivity distribution becomes close to that corresponding to an initial constant density of bugs, formally obtainable for $2r_{0}/L = 1$.

%
\begin{figure}[ht!]
    \centering
    \includegraphics[scale=0.45,angle=0]{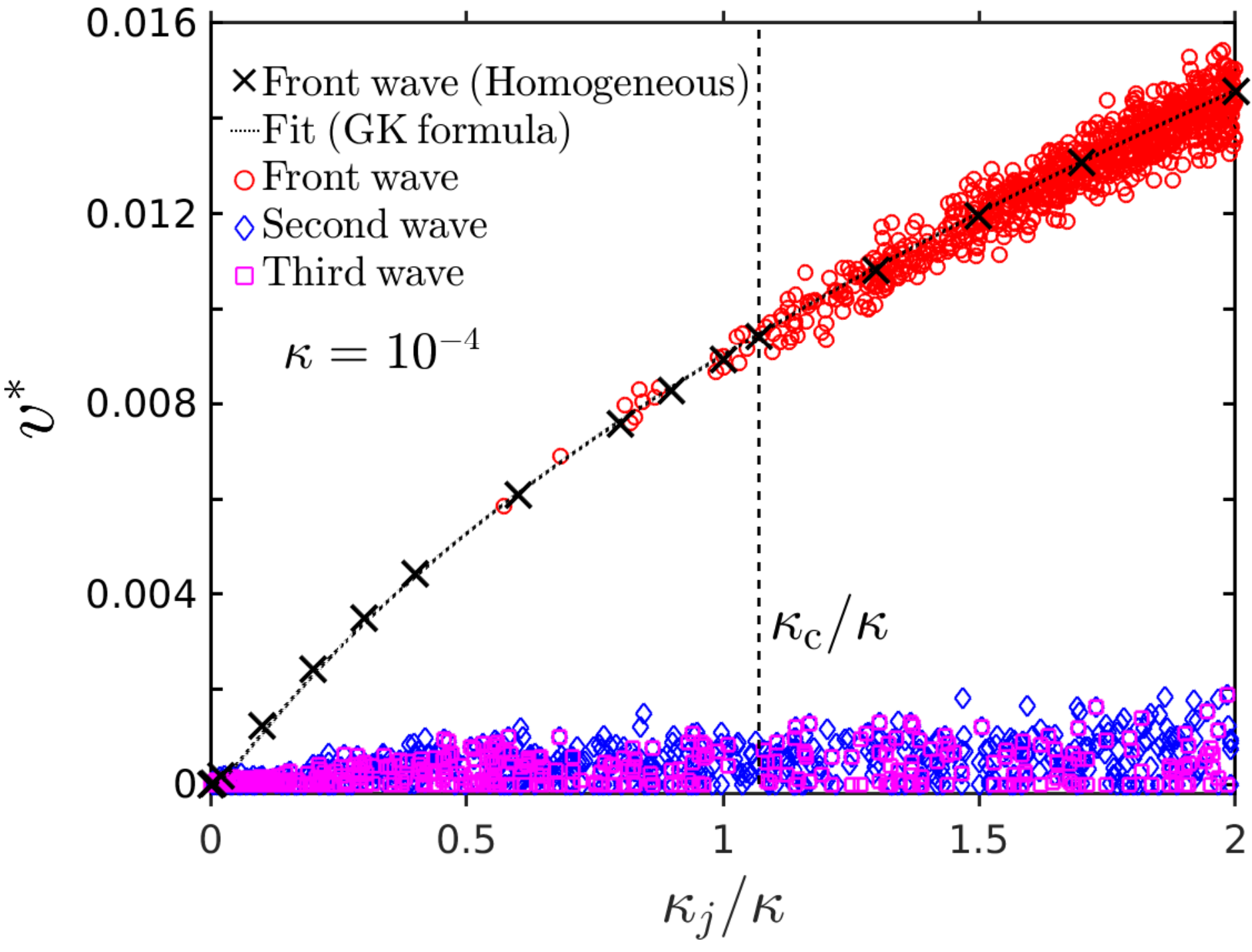}
    \caption{
        Spreading velocities $v^*$ versus $\kappa_j$.
        Black crosses ($\boldsymbol{\times}$): average front wave velocity in the \textit{homogeneous} bugs model (computed over 100 realizations): in this case the abscissa $\kappa_j$ represents the fixed value of diffusivity of all the bugs.
        Dotted line ($\cdot$$\cdot$$\cdot$): fit based on the Ganan-Kessler velocity formula given by Eq.~\eqref{eq_VGK}, with parameters $a \approx 1.46$ and $b \approx 0.3$. 
        Red circles (\textcolor{red}{$\boldsymbol{\Circle}$}): front wave velocities in the \textit{heterogeneous} bugs model (their average values coincide with the Ganan-Kessler velocities).
        Blue diamonds (\textcolor{blue}{$\boldsymbol{\Diamond}$}): ``second wave'' velocities observed in the \textit{heterogeneous} bugs model. 
        Magenta squares (\textcolor{magenta}{$\boldsymbol{\Square}$}): ``third wave'' velocities observed in the \textit{heterogeneous} bugs model.
        All the velocities in the heterogeneous bugs model were computed from single realizations with mean diffusivity $\kappa=10^{-4}$ (400 data overall).
        See text for details.
    }
    \label{fig_08}
\end{figure}
%
\subsection{Front wave velocity}
\label{sec_Gk_vel}

Here, we focus on the spreading process and the corresponding front waves observed in the simulations discussed above.
We will limit ourselves to consider the one-dimensional configuration, both for computational convenience and for avoiding the more complex topology~\cite{Brazhnik-2000} and ecological dynamics~\cite{Malley_et_al_2009} of multi-dimensional systems, focus on the most interesting case $\kappa=10^{-4}$.
As we will discuss below, the study of the velocity associated with the spreading of a species across the domain allows to distinguish between species carrying out the colonization process of a previously empty area from the other species spreading across an already populated area.

In the numerical simulations of the one-dimensional dispersal-structured bugs model, one usually observes first an expansion starting from the initial spot and proceeding in the two opposite directions with well defined wave fronts, usually associated with two different diffusivities, as in the realizations shown in Fig.~\ref{fig_05}.
Later on, other species may start spreading across the domain.

The first phase of the time evolution of the system, in which the colonization process of the empty part of the domain takes place, is dominated on each side by a single colonizing species, even if there can be in the meantime various competing species localized around the initial spot.
For this reason, it is natural to compare the colonization process in a dispersal-structured population due to bugs with a certain diffusivity $\kappa_j$ with the analogous process in a homogeneous population of bugs in which all individuals have the same given diffusivity $\kappa_j$.

In the limit of local interaction ($R \to 0$), the homogeneous bugs model is described in the mean field limit by the Fisher-Kolmogorov-Petrovsky-Piskunov (FKPP) equation~\cite{Fisher37,Murray-2003a}, $\partial_t \, \rho = \kappa \, \partial_x^2 \, \rho + \Delta_0 \rho \, (1 - \rho/\rho^*)$, where $\rho^*$ represents the carrying capacity.
The asymptotic front wave velocity can be derived analytically and is given by the Fisher velocity, $v_{\mathrm{F}}(\Delta_{0},\kappa) = 2\sqrt{\Delta_{0}\kappa}$, when the population is initially bounded within a given region~\cite{Berestycki_2009,Saarloos-2003}.

For the homogeneous bugs model ($R > 0$), described by a non-local FKPP equation, an analogous result valid in the mean field limit is not available.
However, analytical and numerical studies have provided information about the dependence of the front wave velocity on the main system parameters.
Ganan and Kessler studied in Ref.~\cite{Ganan-kessler-2018} the non-local FKPP equation with a Gaussian and flat-top competition kernel.
For the latter type of kernel, corresponding to the homogeneous version of the model studied in this paper, it was found that the Fisher formula for the front wave velocity can be generalized to
\begin{equation}
    v_{\mathrm{GK}}
    =a \sqrt{\Delta_{0}\,\kappa \,} \exp \left(- b \sqrt{\frac{\Delta_{0}}{\kappa N_{R}^{*}}}R \right) \,  ,
    \label{eq_VGK}
\end{equation}
%
where $a$ and $b$ are fitting parameters and the prefactor of the exponential function is proportional to the Fisher velocity.

We measured numerically the front wave velocity in the \textit{homogeneous} bugs model for different values of the diffusivity, using the same parameter values ($\Delta_0=0.9$, $R=0.1$, and $N_{R}^{*}=45$), an initial spot diameter $2r_{0} = R$, and a system size $L=50$ that allows a reliable velocity measurement.
In each realization we computed for each (left and right) colonization process the corresponding front velocity as $v^{*} = x _{\mathrm{front}}(t^{*}) / t^{*}$.
The time $t^{*}=1000$ is long enough to obtain the asymptotic velocity value, but not too long that the wave front had already reached the boundary of the domain;
the position of the wave front $x_{\mathrm{front}}(t)$ at time $t$ was obtained using the tracing method introduced in Ref.~\cite{Saarloos-2003}, i.e., (e.g. for the right wave front) determining the rightmost point of the particle density $\rho(x,t)$ that fulfills the condition $\rho(x_{\mathrm{front}},t) = \rho_{\mathrm{th}}$, where $\rho_{\mathrm{th}}$ is a threshold value assumed to be $\rho_{\mathrm{th}} = \frac{1}{2}\max\,[\rho(x,t^{*})]$.
Then we averaged the front wave velocities thus obtained over 100 different realizations, obtaining the data presented in Fig.~\ref{fig_08} as crosses (``homogeneous front wave''; the abscissa ``$\kappa_j$'' in this case is to be understood as the fixed diffusivity of the homogeneous population). 
The results are in good agreement with the Ganan-Kessler formula, Eq.~\eqref{eq_VGK}, represented by the dotted line.

Next, we analyzed the case of a dispersal-structured population with an initial mean diffusivity $\kappa = 10^{-4}$.
We applied the same method to the front wave velocity associated with the colonization processes observed in the single realizations (without taking averages).
Results for the front wave velocities (400 realizations overall) are plotted in Fig.~\ref{fig_08} as red circles (``Front wave'') versus the diffusivity $\kappa_j$ of the invading species.
Points fall around the curve given by Eq.~\eqref{eq_VGK} and their averages for a given $\kappa_j$ coincide with the Ganan-Kessler curve.

In all the realizations, we observed a diversity $2\le D \le 6$ at time $t^*=1000$, i.e. besides the two invading species (on the left and right side), there are no more than two additional different species on each side.
After computing the wave front velocities on the left and right side, for each realization we repeated the spreading velocity measurement using the same method -- this time not considering the species associated with the wave front.
The velocities obtained in this way are represented as blue diamonds (``second wave'').
In turn, ignoring also the presence of the species associated with the ``second wave'', it is possible to measure the velocity of the remaining species, obtaining the points represented as magenta squares in Fig.~\ref{fig_08} (``third wave'').
The data in Fig.~\ref{fig_08} show that only the bugs producing the wave fronts have velocities described by the Ganan-Kessler formula.
Instead, the velocities computed for the second and third waves are similar to each other, having much lower velocities and not exhibiting any particular velocity pattern.
The reason of this difference is that bugs spread across a domain already populated by other bugs according to mechanisms and time scales different from that characterizing a wave front: 
their spreading across space is more similar to a diffusion process, in which the mean velocity decreases with time, and furthermore the time scale can be much longer, depending on their ability to invade already populated areas.
In other words, the first wave is an actual wave associated with a colonization process.
Instead, in the case of the second and third waves, the term ``wave'' should be understood just as a formal label, since they represent invasion processes taking place as other species diffuse and replace the species present previously --- in this case there is no definite $v^*$-$\kappa_j$ relation and the corresponding points in Fig.~\ref{fig_08} are clustered near the $\kappa_j$-axis.
As an example of such invasion process, one can look at the diffusion of the bugs in Fig.~\ref{fig_05}-(b):
bugs with diffusivities $\kappa_j/\kappa \approx 1$ start an invasion on the right hand side of the initial spot, followed by a more successful invasion of bugs with lower diffusivity $\kappa_j/\kappa < 0.5$, which eventually conquer the whole domain.

The problem of multiple invasions of a habitat carried out by two or more species is characterized by different velocities and time scales, see e.g. Ref.~\cite{Li-2018,Elliot_Cornell_2012}.
In fact, in the specific case of the heterogeneous bugs model, we showed in Refs.~\cite{HNMP-2018,HHL-2013} that when clustering occurs the relaxation times to a steady state with diversity $D = 1$ can diverge.
In this respect, the Brownian bugs model provides an additional example of framework in which coexistence between different species is made possible by the spatial (self-induced) clustering of individuals, confirming previous results along the same line \cite{Brigattietal-2009,Brigattietal-2011,Eigentler-Sherratt_2020,Maciel_Martinez-Garcia_2021}.
However, it is to be noticed that the presence of clustering is not always a necessary condition for coexistence.
As discussed in Ref. \cite{EHG-2015}, according to the mean-field description also different species with large diffusion coefficients, not leading to cluster formation, can coexist, until neutral fluctuations eliminate irreversibly one species after another, until finally only one of them survives.


\section{Discussion and conclusion}
\label{Sec-conclusion}


We studied the dynamics of natural selection in a dispersal-structured population of Brownian bugs, initially localized within a small region of the accessible domain, undergoing a diffusion process and competing with each other for resources through a finite-range interaction.
We observed that the localized initial conditions and the consequent spreading process produce specific features on the natural selection process, which were absent in previously studied situations, in which the initial spatial distribution of bugs was homogeneous across the domain.
Such features appear to be qualitatively independent of the boundary conditions used (periodic or reflecting) and of the dimensionality of the domain (one- or two-dimensional).

The overall process observed is complex and strongly stochastic:
while the initial part of the time evolution, when the colonization process of the empty region takes place, is dominated by one or few families of bugs with high diffusivity, the following part of the time evolution may be dominated by different slower bugs that invade the already populated domain.
Correspondingly, the range of time scales involved is wide, as shown previously~\cite{HNMP-2020,HNMP-2018}, since the invasion of a populated area can be a much slower process than the colonization of an empty region.

In the case of bugs initially distributed  randomly across the whole domain, the selection process immediately favors low diffusivities smaller than the critical one, $\kappa_j \ll \kappac$, giving some competitive advantage to higher diffusivities only at higher levels of demographic noise. 
Instead, in the case of a localized initial spatial distribution of bugs, considered here, all the diffusivities --- low, intermediate, and high --- have crucial roles in the selection process: slower bugs are effective in colonizing an already populated area, whereas faster bugs, which would be otherwise disadvantaged, now may gain a permanent advantage in terms of population size during the colonization process, eventually winning the competition process.
The corresponding final diffusivity distributions $P(\kappa_j$) characterizing the overall natural selection process can have very different shapes and even different trends, e.g. favoring higher instead of lower diffusivities.
In particular, the critical diffusivity $\kappac$, which divides the range of diffusivities into the sub-critical diffusivities associated with a spatial periodic pattern from that of super-critical diffusivities that do not lead to pattern formation, appears to be a suitable compromise and plays an important role in the selection process.

The model studied is inherently stochastic and any species among those initially present in the system can win the competition process; at the same time, the final state, described statistically in terms of the diffusivity distribution of the surviving species, shows some clear trends and dependencies on the system parameters.
A systematic  study of different values of the ratios $r_0/L$, between the initial spot size $r_0$ and the domain size $L$, demonstrates explicitly the strong size effects on the natural selection process.
Furthermore, we have shown that for $r_0/L \to 1$ the results continuously reconnect to those previously obtained for a constant initial distribution of bugs.

We also measured numerically the wave front velocity of the colonization process (in the one-dimensional case and showed that it is well described by the formula of Ganan and Kessler~\cite{Ganan-kessler-2018}.
In general, the velocity of expansion of a species was shown to be a good indicator of the underlying dynamics: 
in fact, only the velocities of wave fronts, each wave front being due to a certain species with diffusivity $\kappa_j$, follow the Ganan-Kessler formula, while the spreading velocity of the other bugs present in the system is much smaller, signaling an invasion process of a species gradually replacing other species.



The present study could contribute to clarify some aspects of the dynamics of competition processes taking place at expanding frontiers, such as those associated with spreading and colonization processes across initially empty regions.
In particular, the colonization process observed in the bugs model presents features resembling the fixation of dispersal phenotypes found in microbial range expansion~\cite{Hallatscheketal_2007}.
The bugs model, which does not describe an evolutionary process, clearly shows at the level of natural selection that bugs with high value of diffusivity have a competitive advantage at the expanding wave front.
This general trend is illustrated by the time evolution of the survival diffusivity distribution $P(\kappa_j)$, e.g. for the one-dimensional system shown in Fig. \ref{fig_06}.
As time passes, the contribution of the largest diffusivities to the distribution increases and reaches a maximum at some time, before decreasing thereafter due to the presence of the boundaries (either periodic or reflecting), which force the bugs at the front wave to diffuse across the system and interact with the other slower bugs. 


These features of the bugs model are connected to the question of the dispersal-competition trade-off, reported in various studies showing that higher dispersal abilities come at the cost of a competitive disadvantage~\cite{Cadotteetal_2006,Larocheetal_2016,Nadell-Bassler-2011}. In other words, individuals can be either good colonizers or good competitors, suggesting a negative correlation between dispersal abilities (described at a modeling level by diffusion coefficients) and competition skills (related to growth rates and/or carrying capacities).

In fact, a dispersal-competition trade-off is a most distinctive feature of the bugs model, in spite of the fact that the heterogeneous bugs model assumes that competition is symmetric, i.e., the effect that individuals exert on the reproduction rate of their neighbors is homogeneous for all bugs and independent of their dispersal phenotypes:
previous studies have shown that slower bugs form stronger clusters that provide them with a higher competitive advantage, see Refs.~\cite{HHL-2013,HNMP-2020,HNMP-2018} for details, naturally accounting for a dispersal-competition trade-off through the finite range of the interaction.
Also the present work, which studies explicitly the expansion of an initially confined population, confirms this general trend showing that not only slower bugs are better competitors locally but also that faster bugs are better colonizers:
in particular, notice that no wave fronts were observed for diffusivities smaller than the value $\kappa_j/\kappa_c < 0.5$ (Fig.~\ref{fig_08}), which also represents the upper cutoff of the final survival probability $P(\kappa_j)$ from constant initial conditions~\cite{HNMP-2020} (black solid curve in Fig. \ref{fig_07}).   
The interval $0.5 < \kappa_j/\kappa < 1$ represents an intermediate situation (see Fig. \ref{fig_08}), where only a few front wave velocities are observed, while most of the front waves are associated with diffusivities $\kappa_j > \kappa$.
However, as discussed above, results are more complex than a direct trade-off: also fast bugs, which are good colonizers by definition, can eventually win the competition process, if the system size is large enough (see Fig.~\ref{fig_07}).

Other questions raised in this work deserve further and more detailed investigations.
In particular, numerical simulations of the dynamics of multiple invasions have revealed a complex dynamics --- in particular that a heterogeneous population of bugs can self-organize into spatially periodic clusters, each cluster possibly populated by a different species~\cite{HNMP-2018,HHL-2013}, but a theoretical description of such a heterogeneous dynamics is still missing. 
We also mention that the wave front associated with the colonization process in a two-dimensional system is definitely more complex than in the one-dimensional case considered, where it is localized within a small moving spatial interval (or two intervals on the opposite sides in the case of initially localized distributions of bugs).
In the two-dimensional case, the wave front is characterized by an inhomogeneous corrugated shape, which has a crucial influence on the system evolution.
A possible numerical approach to the study of the wave front in particular and more in general of the heterogeneous dynamics, which takes into account both the fluctuations and the non-local character of the competition process, is to go beyond the \MF\ approximation by employing methods based on stochastic partial differential equations.


\section*{Acknowledgments}
	The authors acknowledge support from the Estonian Research Council through Grant PRG1059. 
	The authors also thank an anonymous referee for useful suggestions about the connection between the results and the ecological dispersal-competition trade-off.

\end{document}